\documentclass[conference]{IEEEtran}
\IEEEoverridecommandlockouts
\usepackage{xcolor}
\usepackage{cite}
\usepackage{amsmath}
\usepackage{graphics} 
\usepackage{epsfig} 
\usepackage{mathptmx} 
\usepackage{times} 
\usepackage{amsmath} 
\usepackage{amssymb}  
\usepackage{algpseudocode}
\usepackage{algorithm}
\usepackage{graphicx}
\usepackage{cuted}
\usepackage{hyperref}
\usepackage{caption}
\usepackage{subcaption}
\begin{document}

\title{TransRx-6G-V2X : Transformer Encoder-Based Deep Neural Receiver For Next Generation of Cellular Vehicular Communications}

\author{\IEEEauthorblockN{Osama Saleem}
\IEEEauthorblockA{\textit{INSA Rouen Normandie, Univ Rouen Normandie} \\
\textit{Université Le Havre Normandie, Normandie Univ}\\
\textit{LITIS UR 4108, F-76000 Rouen, France} \\
osama.saleem@insa-rouen.fr}
\and
\IEEEauthorblockN{Soheyb Ribouh}
\IEEEauthorblockA{\textit{Univ Rouen Normandie, INSA Rouen Normandie } \\
\textit{Université Le Havre Normandie, Normandie Univ}\\
\textit{LITIS UR 4108, F-76000 Rouen, France} \\
soheyb.ribouh@univ-rouen.fr}
\and
\IEEEauthorblockN{Mohammed Alfaqawi}
\IEEEauthorblockA{\textit{VEDECOM}\\
\textit{Versailles, France} \\
mohammed.alfaqawi@vedecom.fr}
\and
\IEEEauthorblockN{Abdelaziz Bensrhair}
\IEEEauthorblockA{\textit{LITIS, INSA} \\
\textit{Rouen, France} \\
abdelaziz.bensrhair@insa-rouen.fr}
\and
\IEEEauthorblockN{Pierre Merdrignac}
\IEEEauthorblockA{\textit{VEDECOM} \\
\textit{Versailles, France} \\
pierre.merdrignac@vedecom.fr}
}

\maketitle

\begin{abstract}
End-to-end wireless communication is new concept expected to be widely used in the physical layer of future wireless communication systems (6G). It involves the substitution of transmitter and receiver block components with a deep neural network (DNN), aiming to enhance the efficiency of data transmission. This will ensure the transition of autonomous vehicles (AVs)  from self-autonomy to full collaborative autonomy, that requires vehicular connectivity with high data throughput and minimal latency. In this article, we propose a novel neural network receiver based on transformer architecture, named TransRx, designed for vehicle-to-network (V2N) communications. The TransRx system replaces conventional receiver block components in traditional communication setups. We evaluated our proposed system across various scenarios using different parameter sets and velocities ranging from 0 to 120 km/h over Urban Macro-cell (UMa) channels as defined by 3GPP. The results demonstrate that TransRx outperforms the state-of-the-art systems, achieving a 3.5dB improvement in convergence to low Bit Error Rate (BER) compared to convolutional neural network (CNN)-based neural receivers, and an 8dB improvement compared to traditional baseline receiver configurations. Furthermore, our proposed system exhibits robust generalization capabilities, making it suitable for deployment in large-scale environments.
\end{abstract}


\section{INTRODUCTION} \label{s1}

The sixth generation (6G) of cellular networks is expected to revolutionize  wireless connectivity and bring significant advancements in communication system design. Technologies such as massive multiple-input multiple-output (MIMO), millimeter-wave (mmWave) communication, and intelligent reflecting surfaces (RIS) are set to boost the capacity and efficiency of wireless networks, enabling high data throughput with low latency, high reliability, and a high level of security against cyber-attacks\cite{a1}. To achieve this, Artificial Intelligence (AI) will be widely explored for various tasks to improve the performance of the PHY layer. This has been widely motivated by the latest declaration from the 3GPP consortium, which has affirmed the integration of AI in wireless communication system design in the upcoming releases \cite{a11}. The benefits promised by 6G will potentially contribute to accelerating the launch of several technologies, where high wireless network connectivity is required such as Cooperative Connected and Autonomous Mobility (CCAM) which enables smart transportation system through the deployment of autonomous vehicles (AVs). AVs are equipped with multiple sensor, such as video cameras, radars and light detection and ranging (LiDAR) systems. These sensors and devices capture  information and share it with  surrounding vehicles and the environment to facilitate tasks with high precision. These cooperative systems are connected in a vehicular network to ensure information sharing through vehicle-to-everything (V2X) communications.

\begin{figure}[]
  \centering
  \includegraphics[width=0.43\textwidth,height=0.2\textheight]{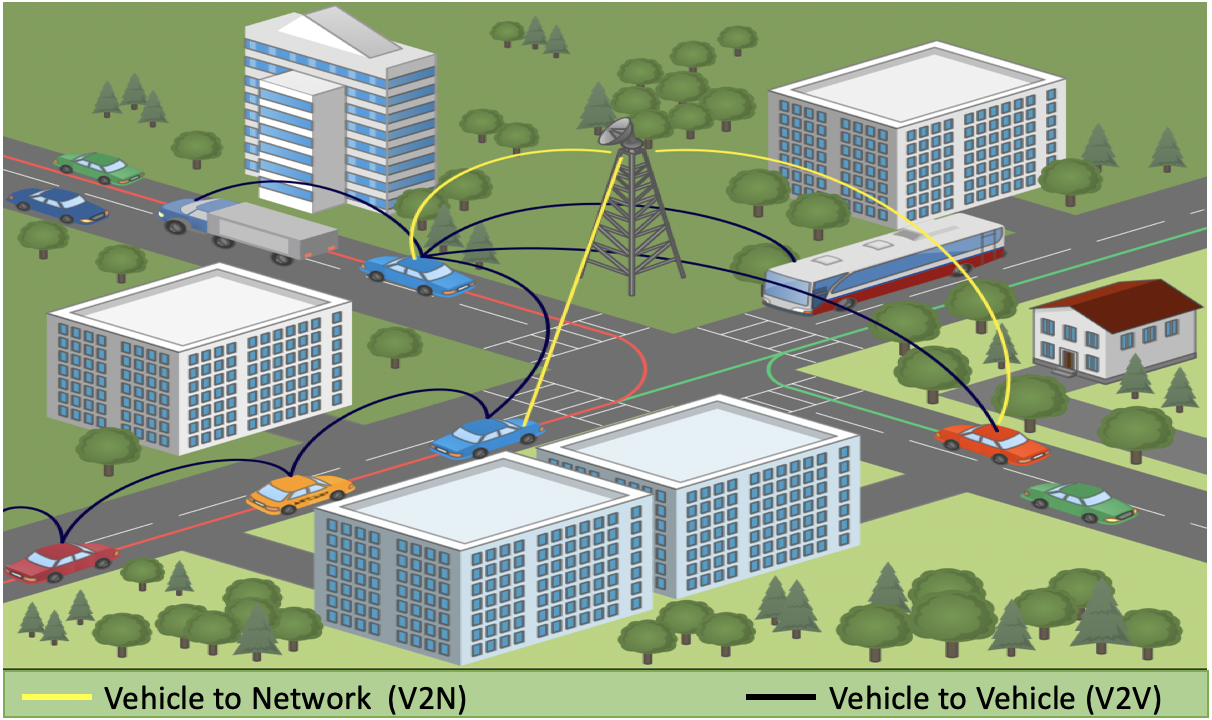}
  \caption{Vehicular Communications in Urban Dense Environment}
  \label{fig1}
\end{figure}

Moreover, applications like remote car control and safe intersection crossing require near real time processing with low error and latency, leading to Ultra-Reliable Low-latency Communication (URLLC) \cite{b1}. The integration of AI into AVs is set to play a pivotal role in maximizing the effectiveness of V2X communication. AI models can be deployed in vehicles to learn from large amounts of real-world data enabling them to make smart decisions. Similarly, the use of Deep Learning (DL) algorithms can further refine the responsiveness and adaptability of V2X communication. DL models can predict and react to changing network conditions, vehicle behaviors, and external environmental factors, thereby sustaining a high level of performance without human intervention. These capabilities are crucial for achieving inflexible quality of service (QoS) demands required by next-generation vehicular networks. 

The concept of auto-encoder-based end-to-end (E2E) wireless communication system was initially proposed in \cite{a2}. The idea was to detect the non-linear imperfections of the real time channel using DL-based approaches. Although the research work in \cite{a2} considered a simple scenario, that has served as a breakthrough for physical layer design in the next generation of wireless communication. Since then, many methods have been proposed for the optimization of transmitter and receiver functionalities using DL. In \cite{c2}, the authors leverages machine learning techniques to estimate conditionally Gaussian random vectors with random covariance matrices, reducing complexity in communication systems by using structured covariance matrices and neural networks to learn efficient estimators. Moreover, He et al., \cite{c3} use an approximation message passing network based on learned denoising to significantly enhance channel estimation in massive MIMO mmWave systems with minimal RF chains, outperforming conventional compressed sensing techniques. Furthermore, Chang et al., \cite{c4} examine how convolutional neural networks (CNNs) are applied to the equalization task. Their methodology yields a reduction in error vector magnitude, surpassing outcomes from established algorithms such as the multi-modulus algorithm and recursive least squares techniques. In the domain of demapping, Shental et al. \cite{c5} propose a deep neural network framework designed to efficiently compute bit Log-Likelihood Ratios (LLRs) for equalized symbols. The efficacy of this deep learning-based demapper closely aligns with that of the optimal log maximum a-posteriori algorithm, albeit with a considerable decrease in computational overhead. Moreover, several methods investigate the integration of deep learning mechanisms into the standard processing sequence of receivers~\cite{c6}\texttt{-}\cite{c8}. These augmented systems demonstrate notable performance improvements over conventional receivers, through proper training. In the above  approaches, although neural networks give close results to the baseline system in terms of performance, where they train their models to perform distinct operations of the receiver independently (e.g., either only estimation, equalization or demodulation).

The joint optimization of receiver functionalities using a neural network has also been addressed by many researchers. Ye et al., \cite{c9} delve into the integration of channel estimation and signal detection via deep learning methodologies. Their approach employs a fully connected neural network to analyze both pilot and data signals, demonstrating significant performance gains over traditional receivers based on minimum mean square error (MMSE) criteria, especially in scenarios with limited channel estimation pilots or in the absence of a cyclic prefix. In contrast, the research work presented in \cite{c10} utilizes convolutional neural networks (CNNs) to construct a receiver capable of deriving bit estimates directly from the time-domain received signal. This methodology exhibits superior performance in low to medium signal-to-noise ratio (SNR) conditions. Although the CNN-based approach continues to outperform receivers based on linear least squares at higher SNRs, it does not achieve the same level of accuracy as MMSE-based receivers with perfect channel knowledge. In~\cite{c11}, the authors propose a deep fully convolutional neural network designed for processing 5th Generation (5G) signals, achieving superior performance over traditional algorithms by leveraging data and pilot symbols for precise channel estimation and generating soft bits compatible with 5G channel coding. Although the research works presented in~\cite{c9}\texttt{-}\cite{c11} propose joint optimization of neural receiver, their works lacked the domain generalization (DG) issue \cite{c12} i.e., when the model is tested on Out-Of-Distribution (OOD) data compared to the training data, its performance degrades drastically. This Domain Generalization (DG) \cite{c12} problem for neural receiver has been addressed in \cite{c13}, where an Urban Micro-cell (UMi) channel model  has been used in training task to avoid overfitting at the time of testing. 

Transformers \cite{c14} have revolutionized the design of AI models, with the attention mechanism outperforming the convolution mechanism due to its ability to capture long range data dependencies in multiple domains \cite{c15}\texttt{-}\cite{c17}. However, their performance has not been evaluated for E2E wireless communication systems. Furthermore, almost all of the proposed neural receivers for wireless communication systems in the literature consider low mobility users in their experiments. In this paper, we propose a novel transformer encoder architecture for neural receiver. This receiver is designed to process 6G data symbols in the uplink scenario of Cellular Vehicle to Network (C-V2N) by handling frequency domain signals and predicting LLRs from the received resource grid. To the best of our knowledge, this is the first work aimed at using an attention-based neural receiver for 6G communication in C-V2N.

\begin{figure*}[t]
    \centering
    \includegraphics[width=0.9\textwidth,height=0.26\textheight]{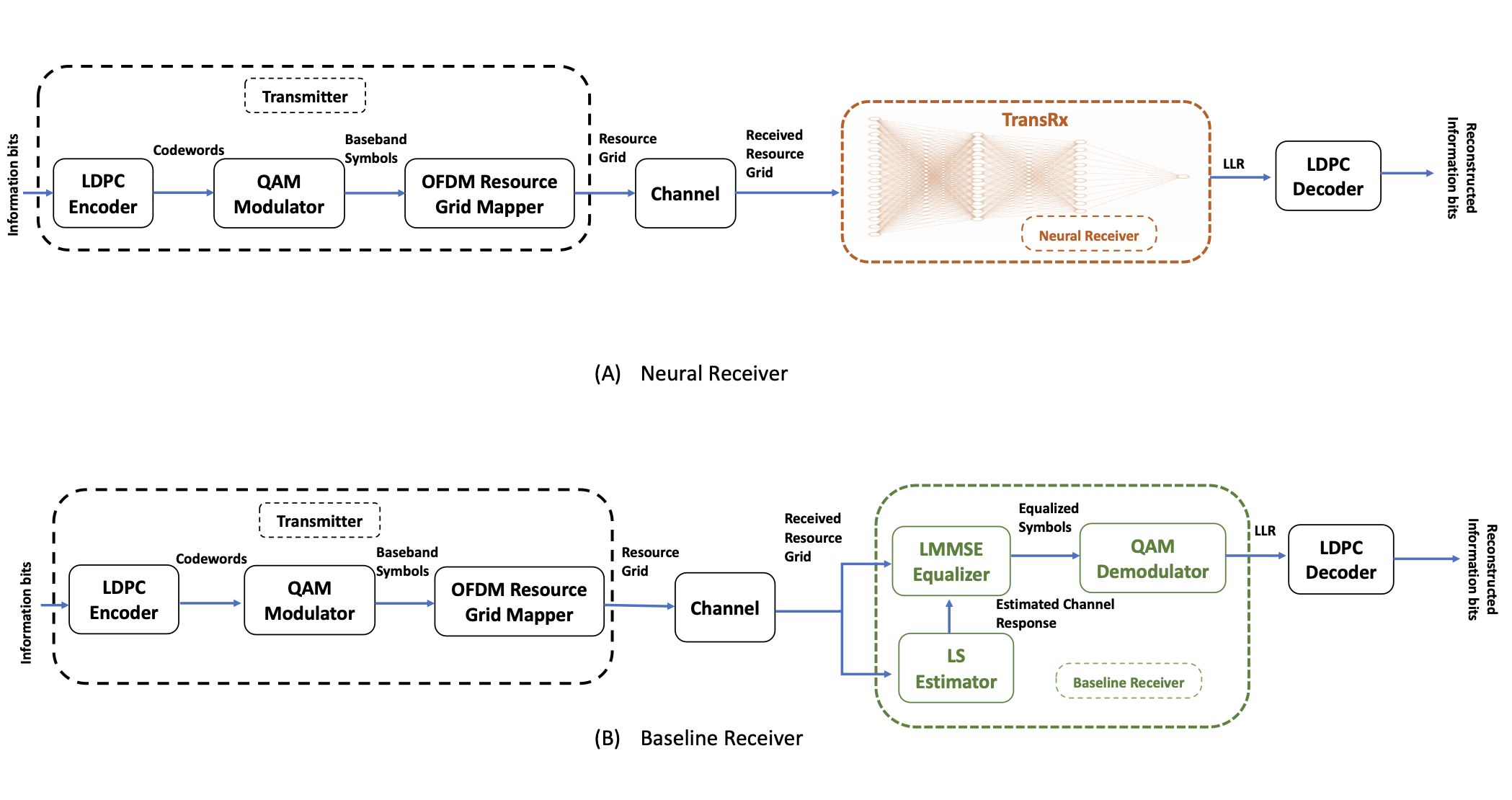}
    \caption{End to end wireless communication}
    \label{fig2}
\end{figure*}

Our main contributions are as follows:
\begin{enumerate}
\item We propose a new deep neural receiver named TransRx designed for 6G-V2X communications. Built on an attention mechanism, TransRx utilizes multiple transformer encoder blocks as its core components to shape the neural network architecture. Our TransRx-6G-V2X receiver takes the received data as input and calculates the optimal LLRs at the output, leading to minimize bit error rate.
\item Our proposed TransRx was tested across various  wireless channel characteristics and scenarios, different from those used in the training task, to evaluate its performance on out-of-distribution data, demonstrating its strong generalization capabilities.

\end{enumerate}
The rest of the paper is organized as follows: In Section~\ref{s3}, we present our proposed neural receiver-based end-to-end communication system. Section \ref{s4} describes our experimental setup scenarios, followed by the results and discussion in Section \ref{s5}. Finally, in Section \ref{s6}, we conclude the paper and outline future work.

\section{System Model} \label{s3}
We consider a set of vehicles $V=\{V_1,V_2,\dots,V_z\}$ operating in an urban dense environment where each vehicle $V_a \in V$, $a = \{1,2, \dots, z\}$ share its information with the base station. As the information bits reach the physical layer of the vehicle, they are first encoded by Low-Density Parity-Check (LDPC) code, as shown in Fig.~\ref{fig2}. This encoded information then goes through the QAM modulator to generate the baseband symbol as output. In the next step, these baseband symbols go through resource grid mapper where known pilot symbols are added for estimating the wireless channel state information at the receiver side. Furthermore, in order to equalize the inter-symbol interference (ISI) at the base station, a cyclic prefix is also added to these symbols. The resulted information is transmitted by the vehicle through the wireless channel. As illustrated in Fig.~\ref{fig2}, once the signal reaches the base station, our proposed TransRx shifts it to log likelihood ratio (LLR) using a transformer encoder-based architecture. For comparison, we also consider the baseline receiver architecture that utilizes the combination of LS Estimator and LMMSE equalizer to recover the original message. Both of these receiver architectures are explained below:


\subsection{Baseline Receiver Architecture} \label{s31}
The received signal at the base station can be expressed as follows:

\begin{equation}
y_{i,j} = H_{i,j} x_{i,j} + n_{i,j} \label{eq1}    
\end{equation}

where $i$ denotes the OFDM symbol and $j$ denotes the subcarrier index. $y_{i,j}$ denotes the received symbols and $x_{i,j}$ denotes the transmitted symbols. $H_{i,j}$ is the channel for $i^{th}$ OFDM symbol and $j^{th}$ subcarrier and $n_{i,j}$ is the channel noise.

As a first step, the base station estimates the channel state information through the known pilot symbols transmitted by the vehicle. The  estimated channel $\hat{H}$ can be computed from eq.~\ref{eq1} as:
\begin{equation}
\hat{H} = y_{i,j} x_{i,j}^* \label{eq2}
\end{equation}

where ($^*$) denotes the complex conjugate. The signal plus interference noise \(\hat{\sigma}^2\) is also estimated at the time of channel estimation. Each of the data symbol received goes through LMMSE equalizer to provide equalized received symbols $\hat{y}_{i,j}$ as follows:
\begin{equation}
\hat{y}_{i,j} = (\hat{H}_{i,j}^H   \hat{H}_{i,j} + \hat{\sigma}^2  I)^{-1} \hat{H}_{i,j}^H   y_{i,j} \label{eq3}
\end{equation}

where ($^H$) denotes the hermition transpose operation, '$I$' denotes the identity matrix and \(\hat{\sigma}^2\) denotes the estimated noise power. The equalized symbols are then shifted to Log likelihood ratio (LLR) by the QAM demmaper as follows:
\begin{equation}
LLR_{i,j,k} \equiv log \frac{P(x_k = 0|\hat{y}_{i,j})}{P(x_k = 1|\hat{y}_{i,j})}    \label{eq4}
\end{equation}

where, $P(x_k = 0|\hat{y}_{i,j})$ is the probability that transmitted bit $x_k$ is '0' given the approximated received symbol $\hat{y}_{i,j}$ and $P(x_k = 1|\hat{y}_{i,j})$ is the probability that transmitted bit $x_k$ is '1' given the approximated received symbol $\hat{y}_{i,j}$. However, $k = 0,1,2,\dots, N-1$ where $N$ denotes the number of bits per symbol. LLRs go through the LDPC decoder to recover the original message $(x_{i,j})$ transmitted by the vehicle as shown in Fig.~\ref{fig2}.

\subsection{Neural Receiver Architecture} \label{s32}

\begin{figure}[t]
  \centering
  \includegraphics[width=0.48\textwidth,height=0.16\textheight]{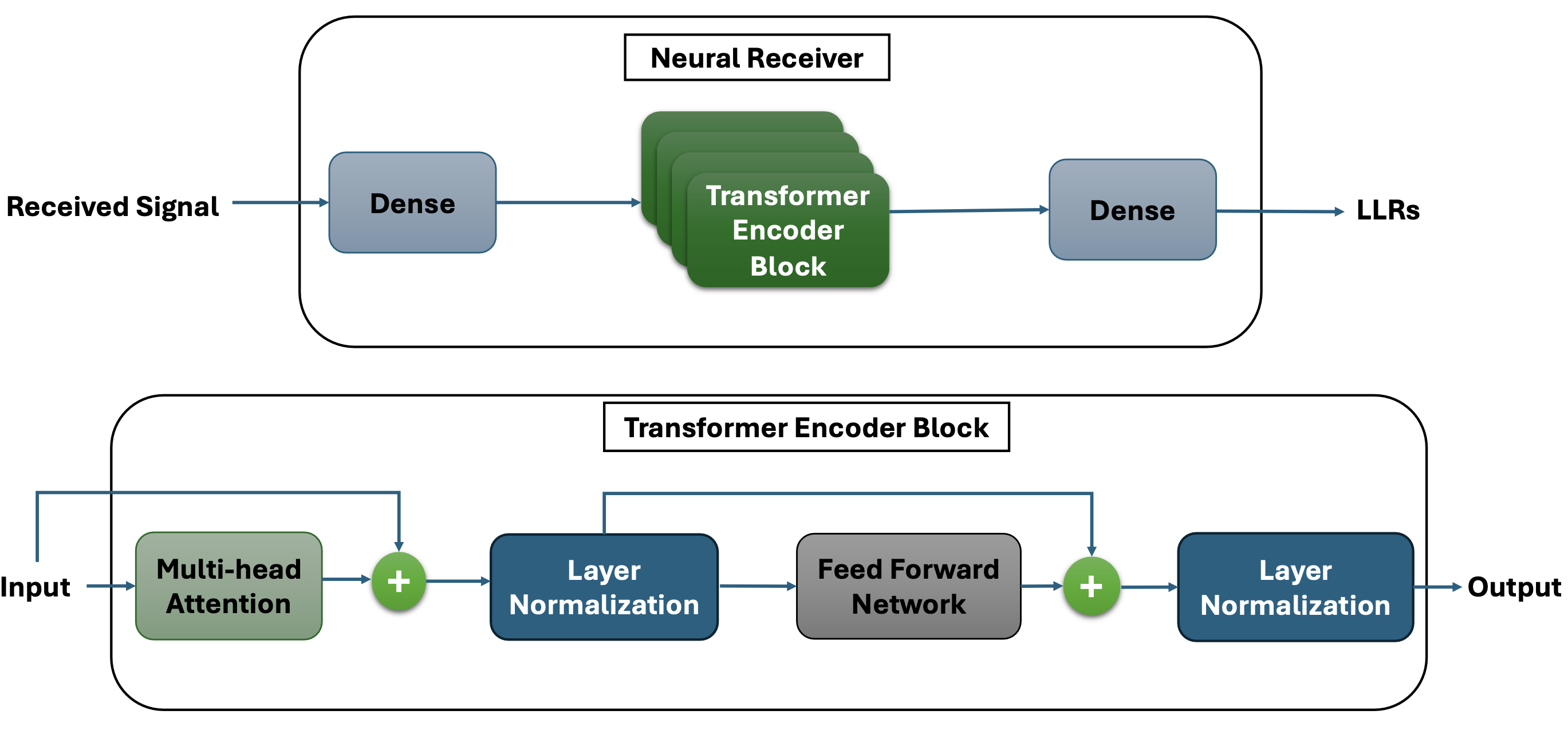}
  \caption{TransRx Architecture}
  \label{fig3}
\end{figure}

Our proposed TransRx, based on  transformer encoder shifts the received signal from the frequency domain to LLRs and replaces the LS estimator, LMMSE equalizer, and QAM demodulator blocks in the baseline receiver. The received signal through the wireless channel, serves as input for TransRx, as shown in Fig.~\ref{fig2}. The TransRx-based neural receiver architecture is illustrated in Fig.~\ref{fig3}. It begins with a dense layer with an output feature dimension of $128$. This is followed by $4$ transformer encoder blocks, each including a multi-head self-attention layer with $4$  attention heads and an embedding dimension of $128$. The output of the multi-head self-attention block is added to the input and then fed into a normalization layer. The output of the normalization layer is subsequently fed into a feed forward network, including $2$ dense layers. Both  of these dense layers have an output feature dimension of $128$, where  the ’relu’ activation function  is applied to the first dense layer. As shown in Fig.~\ref{fig3}, the output of the first normalization layer is added to the output of the feed forward network and fed into a second normalization layer. The final layer of our proposed TransRx architecture is a dense layer with an output feature dimension equal to the number of bits per symbol.

To train the proposed TransRx neural receiver model, we compute bit wise binary cross entropy (BCE) along with the sigmoid activation function to obtain difference between the actual bit value and predicted LLR as shown in Algorithm \ref{algo1}. The BCE can be formulated as:

\begin{equation}
BCE = - \frac{1}{N} \cdot \sum_{i=1}^n  y_i \cdot \log(P(y_i)) + (1 - y_i) \cdot \log(1 - P(y_i)) \label{eq5}     
\end{equation}

where $N$ is the batch size, $n$ is the number of bits in one batch, $y_i$ is the transmitted information bit to be '$1$' and $P(y_i)$ is the probability of LLR to be '$1$'. Similarly, $(1 - y_i)$ is the transmitted information bit to be '$0$', and $(1 - P(y_i))$ is the probability of LLR to be '$0$'. The loss function used to train and update the weights of TransRx is given as :

\begin{equation}
Loss = 1 - \frac{BCE} {\ln{(2)}} \label{eq6}
\end{equation}

\begin{algorithm}[t] 
\caption{TransRx Training}
\label{algo1}
\begin{algorithmic}[1]
\State \textbf{Input}: Initial weights \(\theta^i\), transmitted signal \(x\), received signal \(y\), noise power \(n\)
\State \(y \gets \text{stack}(\text{real[y]}, \text{imaginary[y]})\)
\State \(y \gets \text{concatenate}(y, n)\)
\State \(y \gets \text{Dense}(y)\)
\For{\_ in Transformer\_Layers}
	\State \(y \gets \text{transformer\_layer}(y)\)
\EndFor
\State \(y \gets \text{Dense}(y)\) 
\State \textbf{Error} \( = \text{BCE}(x, y)\)
\State \textbf{Loss} \( = 1 - \text{Error}\)
\State \textbf{Compute: } gradient
\State \textbf{Compute: } weights for next iteration
\State \textbf{Update: } \(\theta\)
\end{algorithmic}
\end{algorithm}

\begin{algorithm}[t]
\caption{TransRx Testing}
\label{algo2}
\begin{algorithmic}[1]
\State \textbf{Input}: Binary data \(x\), noise power \(n\)
\State \(z \gets \text{LDPC Encoder}(x)\)
\State \(z \gets \text{QAM Mapper}(z)\)
\State \(z \gets \text{Resource Grid Mapper}(z)\)
\State \(y \gets \text{Channel}(z, n)\)
\State \(\text{llr} \gets \text{Neural Receiver}(y, n)\)
\State \(\text{llr} \gets \text{Resource Grid Demapper}(\text{llr})\)
\State \(\hat{x} \gets \text{LDPC Decoder}(\text{llr})\)
\State \textbf{Compute: } BER ($x, \hat{x}$)
\end{algorithmic}
\end{algorithm}

\begin{table}[b]
    \centering
    \caption{Simulation Parameters}
    \begin{tabular}{|c|c|} \hline
        \textbf{Parameter} & \textbf{Value} \\ \hline 
        Carrier Frequency & 28GHz \\ \hline
        Physical Channel & UMa \\ \hline
        Modulation & 64 QAM \\ \hline
        Code rate & 0.5 \\ \hline
        Subcarrier Spacing & 240KHz \\ \hline
        Delay Spread & 266ns \\ \hline
        No. of Transmitter Antenna & 1 \\ \hline
        No. of Receiver Antenna & 2 \\ \hline
        No. of OFDM symbol & 14 \\ \hline
        Fast Fourier Transform Size & 128 \\ \hline
        Minimum Vehicle Speed & 60 km/h \\ \hline
        Maximum Vehicle Speed & 120 km/h \\ \hline
    \end{tabular}
    \label{table1}
\end{table}

\section{Implementation and Experimental Setup} \label{s4}
In our experiments, we consider an uplink scenario, where information is sent from the vehicle to the base station in an urban environment through an Urban Macro-cell (UMa) channel specified by 3GPP \cite{c19}. We assume that the vehicle speed ranges from 60 to 120km/h. As described in Algorithm~\ref{algo2}, the message is encoded and modulated before transmission. Once the information is transmitted by the vehicle, it goes through an urban wireless channel. As the message is received by the base station (BS), it is transformed into the LLRs using our proposed TransRx neural receiver. These LLRs are shifted to the approximated original message (in the form of binary signal) sent by the vehicle through LDPC Decoder. Our proposed TransRx neural receiver has been implemented using the open source library for physical layer research called sionna \cite{c20}. TransRx is trained on Tesla P40 graphics card having 24GB of GPU with 12 million data. We created a $6$G compatible wireless communication system using simulation parameters shown in Table~\ref{table1}. The values of carrier frequency, QAM modulation scheme, subcarrier spacing   and coderate are set to $28$GHz, $64$,  $240$KHz and $0.5$ respectively. The number of the antenna transmission in the vehicle is  $1$, while the number of antenna receiver at the base station is $2$. 
\begin{table}[b]
    \centering
    \caption{TransRx  Parameters}
    \begin{tabular}{|c|c|} \hline
        \textbf{Parameter} & \textbf{Value} \\ \hline
        No. of Transformer Blocks & 4 \\ \hline
        No. of Attention heads & 4 \\ \hline
        Feed Forward Network Dimension & 128 \\ \hline
        Embedding Dimension & 128 \\ \hline
        Learning Rate & 1$e^{-3}$  \\ \hline
        Optimizer & AdamW \\ \hline
        Activation Function & Relu \\ \hline
        Training data & 12M \\ \hline
    \end{tabular}  
    \label{table2}
\end{table}

\section{Results and Discussion} \label{s5}
In this section, we present the experimental results achieved using our proposed TransRx. These results have been compared to the following state-of-the-art schemes:

\begin{itemize}
    \item \textbf{Perfect CSI}, this knows perfect channel state information (CSI) with zero error Variance.
    \item \textbf{LS Estimator}, this estimates the channel conditions based on data symbols and subcarriers.
    \item \textbf{DeepRx}, this CNN-based neural receiver architecture has been sourced from \cite{c11}. For a fair comparison with the proposed TransRx neural receiver, we re-trained the CNN model using the same data and UMa channel model. 
\end{itemize}

\begin{figure}[htb!]
  \centering
  \includegraphics[width=0.45\textwidth,height=0.25\textheight]{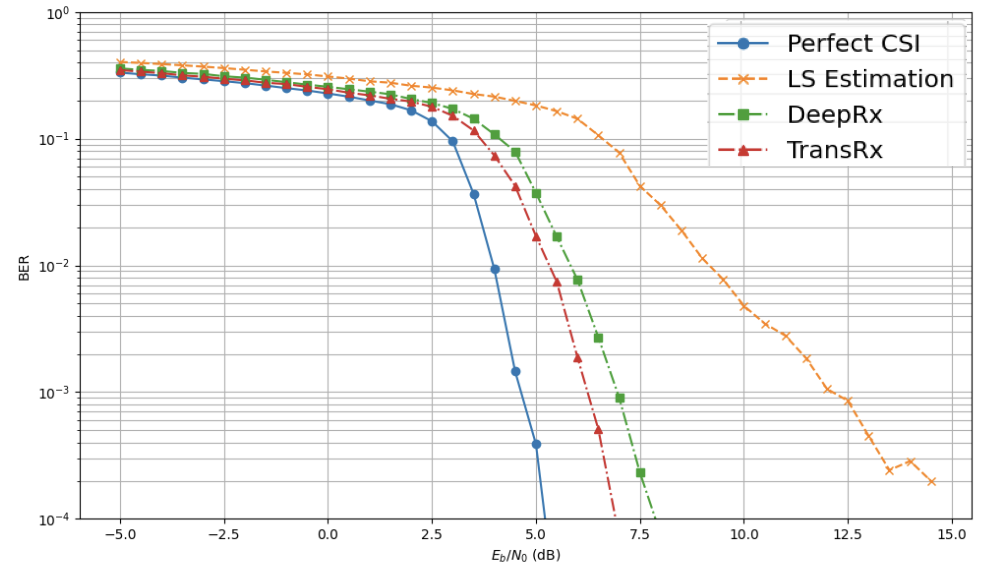}
  \caption{Comparison of BER w.r.t SNR for UMa Channel Model with vehicle speed ranging from 60 to 120 km/h}
  \label{fig4}
\end{figure}

The achieved performance of the proposed TransRx compared to state-of-the-art models are presented in Fig. \ref{fig4}. It shows the BER comparison of the different receiver architectures with respect to (w.r.t) Signal to noise ratio (SNR) over the UMa channel model. We observe that the proposed TransRx outperforms state-of-the-art methods, including LS estimator and Deep Rx. Our proposed TransRx converges to a BER close to 0 at an SNR value of $6.25$ dB, which is close to that achieved by the perfect CSI. In contrast, Deep Rx and the LS estimator reach a BER close to 0 at higher SNR values, around $7.5$ dB and $15$ dB respectively, showcasing a $1.25$ dB and $11$ dB disparity compared to our TransRx. 


\begin{figure}[htb!]
  \centering
  \includegraphics[width=0.45\textwidth,height=0.25\textheight]{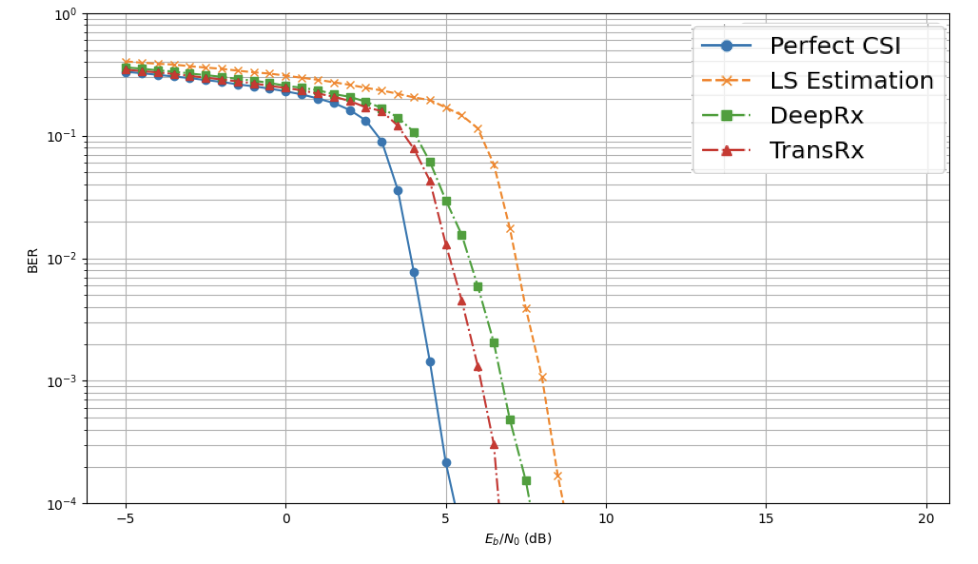}
  \caption{Comparison of BER w.r.t SNR for UMa Channel Model with vehicle speed ranging from 0 to 60 km/h}
  \label{fig5}
\end{figure}

\begin{figure}[htb!]
  \centering
  \includegraphics[width=0.45\textwidth,height=0.25\textheight]{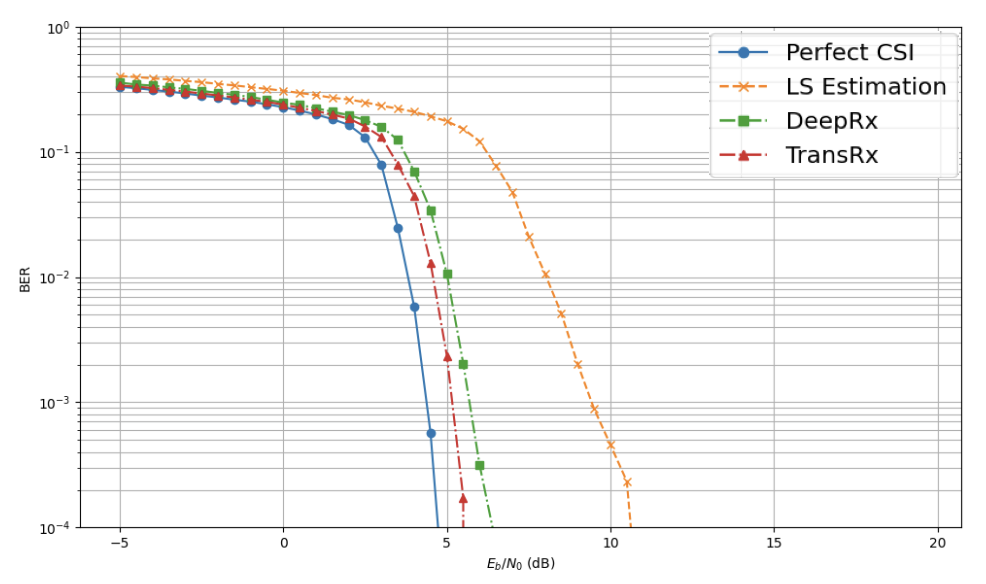}
  \caption{Comparison of BER w.r.t SNR for CDL Channel Model with vehicle speed ranging from 60 to 120 km/h}
  \label{fig6}
\end{figure}

\begin{figure*}[htbp]
    \centering
    \begin{subfigure}{0.24\textwidth}
        \centering
        \includegraphics[width=\textwidth]{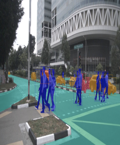}
        \caption{}
        \label{fig71}
    \end{subfigure}%
    \begin{subfigure}{0.24\textwidth}
        \centering
        \includegraphics[width=\textwidth]{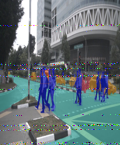}
        \caption{}
        \label{fig72}
    \end{subfigure}%
    \begin{subfigure}{0.24\textwidth}
        \centering
        \includegraphics[width=\textwidth]{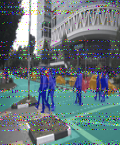}
        \caption{}
        \label{fig73}
    \end{subfigure}%
    \begin{subfigure}{0.24\textwidth}
        \centering
        \includegraphics[width=\textwidth]{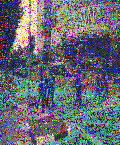}
        \caption{}
        \label{fig74}
    \end{subfigure}
    \caption{(a) Image transmitted by vehicle (b) Image reconstructed at the base station by TransRx (c) Image reconstructed at the base station by DeepRx (d) Image reconstructed at the base station by LS Estimator}
    \label{fig7}
\end{figure*}

\begin{table}[t]
    \centering
    \caption{PSNR Comparison}
    \begin{tabular}{|c|c|} \hline
        \textbf{Algorithm} & \textbf{PSNR (dB)} \\ \hline
        TransRx & \textbf{36.21} \\ \hline
        DeepRx & 34.05 \\ \hline
        LS Estimator & 29.66 \\ \hline
    \end{tabular}  
    \label{table3}
\end{table}

To validate the proposed TransRx on a large scale, we evaluated our neural receiver in different scenarios beyond the one it was trained on. Given that the vehicle operates in a dense urban environment, it is likely that the vehicle speed will reduce at certain instances. We tested the performance of TransRx at vehicle speeds ranging from 0 km/h to 60 km/h. The performance evaluation of the BER over different values of SNR is shown in Fig.~\ref{fig5}. We can see that TransRx still performs well and outperforms state-of-the-art models, with minimal BER improvements of approximately $1$ dB and $2.5$ dB compared to DeepRx and the LS estimator, respectively.


In addition, since the vehicular channel in a real-time environment frequently changes due to high mobility, we evaluated the performance of TransRx over a Clustered Delay Line (CDL) channel with vehicle speeds ranging from 60 to 120 km/h. As shown in Fig.~\ref{fig6}, the performance of TransRx in terms of the achieved BER under various SNR values is superior to state-of-the-art methods, achieving a low BER with an SNR disparity of $3.5$ dB compared to DeepRx and $8$ dB compared to the LS estimator.


We assess the performance of our proposed TransRx for image transmission use cases. We used a real-world image taken by an AV as information shared from the vehicle to the base station at a speed of 90 km/h over an UMa channel model at an SNR of 5 dB. The reconstructed images at the receiver side using TransRx and the state-of-the-art methods are illustrated in Fig.\ref{fig7}. Compared to the benchmark receiver algorithms, our proposed TransRx can reconstruct images with fewer missing pixels, resulting in an image closer to the one transmitted by the vehicle. To quantify the performance of TransRx, we compute the Peak Signal-to-Noise Ratio (PSNR) by measuring the reconstruction quality of the received image compared to the transmitted one. The obtained PSNR for TransRx and the state-of-the-art approaches is shown in Table.\ref{table3}. It is clear that TransRx achieves the highest PSNR (\textbf{36.21dB}), outperforming DeepRx and the LS estimator, which achieve $34.05$ dB and $29.66$ dB, respectively. The success of the TransRx receiver can be attributed to the attention mechanism's ability to capture long-range data dependencies compared to convolution.


\section{Conclusion and Future Work} \label{s6}
In this paper, we proposed an attention-based TransRx architecture as a neural receiver in the C-V2N uplink scenario. The proposed TransRx neural receiver extracts LLR from the frequency domain signal received at the base station. Simulation result shows that once trained on appropriate data, TransRx outperforms the state of the art (DeepRx) neural receiver model and LS estimator in all scenarios. Our proposed TransRx neural receiver shows $3.5$dB improvement to achieve minimal BER compared to the state of the art DeepRx and $11$dB improvement to achieve the same compared to the LS estimator. As a future direction, we are planning to have a testbed implementation of the proposed TransRx-6G-V2X based neural receiver.

\end{document}